\documentclass[prl,aps,showpacs,twocolumn]{revtex4}
\usepackage{graphicx}
\usepackage{amsmath}
\usepackage{amsfonts}
\usepackage{amssymb}

\def\cut#1{}%

\def\MT{_{\rm MT}}
\def\off{^{\rm off}}
\def\on{^{\rm on}}
\def\ssw{_{\rm sw}}
\def\smax{_{\rm max}}
\def\opt{^{\rm opt}}
\def\mot{_{\rm mot}}
\def\eff{_{\rm eff}}
\def\sth{_{\rm th}}
\def\micron{\mu{\rm m}}
\def\second{{\rm s}}
\def\pascal{{\rm Pa}}
\def\eqr{Eq.$\,$}
\def\figr{Fig.$\,$}

\newcommand{\ie}{\textit{i.e., }}

\begin{document}

\title{Hitchhiking Through the Cytoplasm}
\author{Igor M. Kuli\'{c}}

\email[\vspace{-2mm}E-mail: ]{kulic@sas.upenn.edu}
\author{Philip C. Nelson}

\affiliation{Department of Physics and Astronomy, University of
Pennsylvania Philadelphia, PA 19104, USA}

\date{\today}
\begin{abstract}
We propose an alternative mechanism for intracellular cargo
transport which results from motor induced longitudinal
fluctuations of cytoskeletal microtubules (MT). The longitudinal
fluctuations combined with transient cargo binding to the MTs lead
to long range transport even for cargos and vesicles having no
molecular motors on them. The proposed transport mechanism, which
we call ``hitchhiking'', provides a consistent explanation for the
broadly observed yet still mysterious phenomenon of bidirectional
transport along MTs. We show that cells exploiting the hitchhiking
mechanism can effectively up- and down-regulate the transport of
different vesicles by tuning their binding kinetics to
characteristic MT oscillation frequencies.
\end{abstract}

\maketitle

Molecular motor mediated transport along microtubules (MTs) is a well studied
phenomenon in vitro. There is an increasing number of studies of classic
microtubule motors like kinesin and dynein that shed light on their
mechanochemistry in the idealized situation of in vitro single molecule assays
\cite{Kinesin/Dynein}. Despite significant in vitro advances, understanding
how intracellular transport works in vivo still remains one of the big
challenges in molecular biology. Questions like how cellular cargo vesicles
find their way through the cytoplasm and get targeted to their temporary or
final destinations are at the heart of the problem. One of the major mysteries
in this context is a phenomenon called ``bidirectional
transport'' (BDT)~\cite{BiDirectional,BiDirectional2}: The
majority of cargos in the cell move in a bidirectional and remarkably symmetric
manner. Despite the known kinetic and dynamic asymmetry of the underlying +
and $-$ end directed motors (kinesin and dynein respectively), the vesicles
seem to move with the same rates and run length distributions, and exhibit identical
stalling forces, in each
direction  \cite{BiDirectional,BiDirectional2}.

Previous work on BDT has been dominated by the search for a
hypothesized coupling element coordinating the actions of motors
attached to a cargo vesicle to produce the observed back and forth
symmetric motion. But after years of quest for the molecules that
establish the predicted coupling between dynein and kinesin, the
molecular mechanism of the putative coordination model
\cite{BiDirectional} remains obscure. In its very basis the
coordination model remains controversial, as it is hard to imagine
any molecular regulator that would symmetrize the behavior of two
so distinct motor species like dynein and kinesin. Nevertheless
there is strong evidence from several model organisms
\cite{BiDirectional,BiDirectional2} that when either dynein or
kinesin are impeded in their function the bidirectional transport
gets affected in both + and $-$ directions symmetrically. Here we
propose a physically and biologically plausible molecular model
for bidirectional transport that can account for the symmetric
behavior in both directions.

Our model has two basic ingredients: \textbf{1. Motors of either
or both types (not necessarily localized to the cargo) generate
longitudinal MT fluctuations} by already known mechanisms,
\textit{e.g.} the one depicted in \figr1a. This motor induced
confined MT sliding we call ``jabberwalking'' of the underlying
motors. While inter-MT shearing forces are best known for motile
organelles containing the axoneme structure \cite{Amos}, an
increasing number of studies show vigorous sliding of cytoplasmic
MTs when they become detached from cellular structures \cite{MT
Sliding}. Two-point microrheology studies of cytoskeletal
stress-strain fluctuations that demonstrate the presence active
force doublets \cite{Lau} go along the same line of evidence.
\textbf{2. Vesicles transiently couple to the random,
stochastically oscillating MTs}  and attain their speed
temporarily. As a consequence, the vesicles experience strongly
enhanced diffusion. This process we call ``hitchhiking'' of the
vesicles. The occurrence of such a mechanism is most clearly
observed for reticulopodial cytoskeleton extensions \cite{Orokos}
and chlamydomonas flagella \cite{Bloodgood}. In those systems,
artificial and endogenous cargos (\figr1c) are found to move as a
consequence of motor driven MT sliding in a bidirectional manner.
More recently, evidence was found for hitchhiking in S2 drosophila
cell lines \cite{InPrep} indicating a possible ubiquity of the
process.

\begin{figure}[ptb]
\includegraphics*[width=8cm]{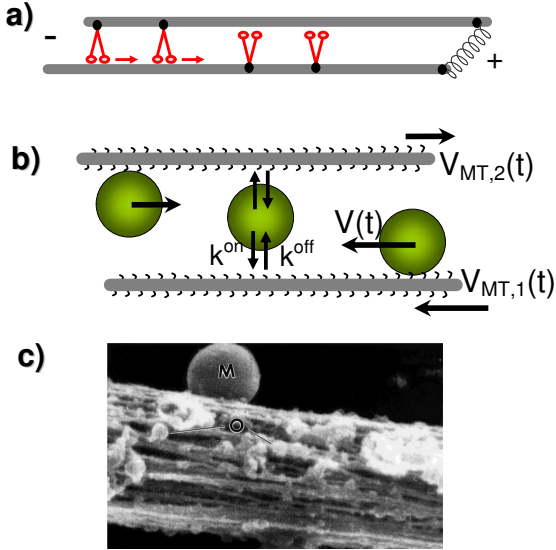}\caption{\textit{a.}$\,$Longitudinal MT noise caused
by ``jabberwalking'' of double attached motors on elastically
tethered MTs. \textit{b.}$\,$The hitchhiking mechanism: Vesicle
binding and unbinding kinetics, combined with local MT
oscillations, induces diffusive, long range transport. MT surface
polarity can also give rise to a directed drift.
\textit{c.}$\,$\textit{Reticulomyxa} transports microspheres (M)
and organelles (O) via MT sliding (adapted from
\cite{Orokos}).}%
\label{Jabberwalking}%
\end{figure}

Below we first focus on the hitchhiking process and compute how a
vesicle transiently coupling to a single longitudinally
oscillating MT moves on long timescales (\figr1b). Each time the
cargo binds to the MT, it assumes the latter's direction and
velocity $v=v\MT $; when it is unbound, we assume complete
immobility \ie  $v=0$. We neglect here the thermal diffusion of
the unbound vesicle due to its smallness compared to the active
transport we consider here (see discussion  below). The velocity
$v\MT \left(  t\right)  $ itself is a random process with the
particular property that $x\MT \left(  t\right) =\int_{0}^{t}v\MT
\left(  \tau\right)  d\tau$ is a bounded variable, \ie  we assume
the MT to move in spatially confined fashion. Consequently a cargo
simply permanently sticking to the MT does not get far, and so
moves in a confined manner as well. However by virtue of the
switching (attachment/detachment) process, the cargo coordinate
$x\left(  t\right)  $ can become unbounded!

The vesicle binding process is described by a 2-state random
variable $B\left(  t\right)  \in\left\{  0,1\right\}  $ with the
characteristics of telegraphic noise \cite{HorsthemkeLefever}:
$B\left(  t\right)  $ equals $1$ if the cargo sticks to the MT or
$0$ for a detached resting cargo. We denote the stochastic
switching rates between those states by $k\off $ and $k\on $. We
next write the velocity of the cargo vesicle as a composite random
process $v\left(  t\right) =B\left(  t\right)  v\MT \left(
t\right)  $. The cargo velocity correlation function is given by
$\left\langle v\left( t_{1}\right)  v\left(  t_{2}\right)
\right\rangle =\left\langle B\left( t_{1}\right)  B\left(
t_{2}\right)  v\MT \left( t_{1}\right)  v\MT \left( t_{2}\right)
\right\rangle $, where $\left\langle \cdots\right\rangle $ denotes
the ensemble average over all realizations of $B$ and $v$. For
simplicity, at first we focus on the limiting case of velocity
independent rates $k\off $ and $k\on $ (we consider the general
case of velocity dependence below). In this case, the vesicle
binding becomes statistically independent\ from the MT motion
$\left\langle Bv\MT \right\rangle =$ $\left\langle B\right\rangle
\left\langle v\MT \right\rangle $, $B\left( t\right)  $ becomes a
standard (asymmetric) Markovian telegraphic noise process
\cite{HorsthemkeLefever}, and we easily compute $\left\langle
v\left(
t_{1}\right)  v\left(  t_{2}\right)  \right\rangle =C\MT \left(  t_{1}%
,t_{2}\right)  C_{B}\left(  t_{1}-t_{2}\right)  $ with $C\MT \left(
t_{1},t_{2}\right)  =\left\langle v\MT \left(  t_{1}\right)  v\MT \left(
t_{2}\right)  \right\rangle $ and $C_{B}\left(  t_{1}-t_{2}\right)
=K^{2}+K\left(  1-K\right)  e^{-2\left|  t_{1}-t_{2}\right|  /t\ssw }.$ The
two constants $K$ and $t\ssw $ characterize the binding behavior.
$K=\left\langle B\right\rangle =k\on /\left(  k\on +k\off \right)  $
denotes the equilibrium binding constant and $t\ssw =2/\left(  k\on %
+k\off \right)  $ is the ``mean switching time.'' Note that whereas the
switching process $B\left(  t\right)  $ is assumed to be stationary
in the
statistical sense, the velocity $v\MT \left(  t\right)  $ can in general be a
non-stationary stochastic or even a purely deterministic process, and its
autocorrelation function $C\MT \left(  t_{1},t_{2}\right)  $ is not
necessarily time homogeneous. The mean-square-displacement (msd) of such a
vesicle is obtained from
\begin{equation}
\left\langle \left(  x\left(  t\right)  -x\left(  0\right)  \right)
^{2}\right\rangle =\int\nolimits_{0}^{t}\int\nolimits_{0}^{t}C\MT \left(
t_{1},t_{2}\right)  C_{B}\left(  t_{1}-t_{2}\right)  dt_{1}dt_{2} \label{MSD}%
\end{equation}
This general expression relates the msd of the cargo to the fluctuations of a
single microtubule and the binding/unbinding kinetics of the cargo to the MT.
For the special case of a statistically stationary process $v_{1}\left(
t\right)  $, the correlator $C\MT \left(  t_{1},t_{2}\right)  =C\MT \left(
t_{1}-t_{2}\right)  $ becomes homogeneous in time and we can simplify
\eqr\ref{MSD} further
\begin{equation}
\left\langle \left(  x\left(  t\right)  -x\left(  0\right)  \right)
^{2}\right\rangle =2\int_{0}^{t}\left(  t-\tau\right)  C_{B}\left(
\tau\right)  C\MT \left(  \tau\right)  d\tau\label{MSDstationary}%
\end{equation}
Now we illustrate our general formulas by focusing on two
particular possible MT shaking processes: (1)~Deterministic
oscillation of the MT with $v_{MT}\left(  t\right)  =V\sin\omega
t$, and (2)~An idealization of MT motion as an overdamped harmonic
oscillator driven by Markovian telegraphic noise (representing the
action of motors, \figr1a).

\textit{Periodically oscillating MT.} In the first case, we have
$C\MT \left( t_{1},t_{2}\right)  =V^{2}\sin\omega t_{1}\sin\omega
t_{2}$, and it is easy to evaluate the rhs of \eqr\ref{MSD}
(\eqr\ref{MSDstationary} cannot be used because of the
non-stationarity of $v_{1}\left(  t\right)  $). The result can be
simplified in two limiting cases. On short timescales $t\ll t\ssw
$, the vesicle does not have enough time to bind/unbind from the
MT, and we simply have $C_{B}\left(  t\right)  \approx K$. From
\eqr\ref{MSD} we obtain $\left\langle \left(  x\left(  t\right)
-x\left(  0\right)  \right)  ^{2}\right\rangle =K\omega^{-2}V^{2}\left(  1-\cos t\omega\right)  ^{2}$, as expected for a
particle strictly following the MT with probability $K$. In
particular, the vesicle stays spatially confined. In the opposite
limit $t\gg t\ssw $ the vesicle motion becomes diffusive:
$\left\langle \left(  x\left(  t\right) -x\left(  0\right) \right)
^{2}\right\rangle ={\rm const.}+2Dt$, with the diffusion constant
given by
\begin{equation}
D=K\left(  1-K\right)  V^{2}%
t\ssw \left(  4+\omega ^{2}t\ssw ^{2}\right)  ^{-1}
\label{DConstantOscillator}%
\end{equation}
This remarkable expression says that a vesicle stochastically
coupling to a periodically oscillating MT diffuses with an
efficiency that depends on the fine tuning of the MT oscillation
frequency $\omega $ and the stochastic vesicle switching time
$t\ssw $. The long time
($t\gg t\ssw$) transport efficiency is maximized for $K\opt =1/2$ and $t\ssw %
\opt =2\omega ^{-1}$, and falls off to zero away from these
values. The intuitive meaning of the first result ($K\opt =1/2$)
is clear: If the vesicle sticks too strongly to the MT ($K=1$) or
not at all ($K=0$), there is no long range transport as it either
moves with the MT (in a confined manner) or not at all.\ The
optimum occurs for an intermediate value. The second finding
$t\ssw \opt =2\omega ^{-1}$, which resembles stochastic resonance
phenomena \cite{StochRes}, also has a simple interpretation: If
the vesicle takes too long a ride on the MT\ ($t\ssw \gg
\omega^{-1}$), its average displacement cancels because of the
pure ``back and forth'' motion of the MT. If the ride is too short
($t\ssw \ll \omega ^{-1}$), a similar argument applies. In this
rather trivial example, we already see an interesting theme:
Optimal transport of vesicles requires a fine tuning of vesicle
binding and MT oscillations.

\textit{MT driven by stochastic motor noise. }In a second more
realistic
approach, we model longitudinal MT oscillations by an overdamped Langevin
equation with a harmonic restoring force coming from
MT\ attachment/confinement (\figr1a,b): $\xi\dot{x}\MT =-Cx\MT %
+F\mot \left(  \dot{x}\MT ,t\right)  .$ Here $\xi$ is the MT longitudinal
friction constant and $C$ the MT restoring spring constant. The actively
generated motor force $F\mot \left(  \dot{x}\MT ,t\right)  $ depends on
the detailed motor mechanochemistry, which we effectively model by a
linear force--velocity relation $v\mot \left(  F\mot \right)  /v_{0}%
=1-F\mot /F_{0}$ with two parameters: $v_{0}$, the maximal (zero-load)
velocity, and $F_{0}$, the motor stalling force. Combining this with the Langevin
equation and $\dot{x}\MT =v\mot $ yields the effective equation of MT
motion:
\begin{equation}
\xi\eff \left(  t\right)  \dot{x}\MT =-Cx\MT +F_{0}\left(  t\right)
\label{LangEff}%
\end{equation}
$\xi\eff \left(  t\right)  =\xi
+F_{0}\left(  t\right)  /v_{0}\left(  t\right)  $ is
the effective friction constant.

\eqr\ref{LangEff} states that the motors contribute to an
effective external force $F_{0}\left( t\right)  $, but also give
rise to increased effective friction $\xi \eff >\xi$. The driving
motor force $F_{0}\left(  t\right)  $ and velocity $v_{0}\left(
t\right)  $ are stochastic variables, which can switch between two
values. The dynamics of this switching generally depends on the
MT--motor attachment geometry. In the simplest arrangement
(\figr1a), motors bind rigidly and run actively on both MTs in a
symmetric manner. In this case $F_{0}\left( t\right) $ and
$v_{0}\left( t\right)  $ both switch between two values, which for
simplicity we assume to be equal in magnitude but of opposite
sign, \ie $F_{0}\left( t\right) =\pm F_{0}$ and $v_{0}\left(
t\right) =\pm v_{0}$ (same number of motors of same strength on
both sides), which results in a time independent friction constant
$\xi\eff =\xi+F_{0}/v_{0}$. We assume that the motors
stochastically switch direction with an exponentially distributed
switching time, \ie $\ F_{0}\left( t\right)  $ is described by
symmetric Markovian telegraph noise \cite{HorsthemkeLefever} with
$p\left( F_{0}\left( t\right)  =\pm F_{0}\right)  =1/2$ and
$\left\langle F_{0}\left( t_{1}\right) F_{0}\left(  t_{2}\right)
\right\rangle ={F_0}^{2}\exp\left( -2\left|  t_{1}-t_{2}\right|
/T_{p}\right) $. Here $T_{p}$ is the processivity time of the
motors (average time between direction changes).

In long time limit, where $x\MT $ in
\eqr\ref{LangEff} becomes a stationary process, we can exploit
\eqr\ref{MSDstationary} to evaluate the msd of the vesicle, provided that we
can compute $C\MT \left(  \tau\right)  =\left\langle \dot{x}\MT \left(
t+\tau\right)  \dot{x}\MT \left(  t\right)  \right\rangle $. To accomplish
this, we use the solution of \eqr\ref{LangEff} in the limit $\xi\eff %
^{-1}Ct\gg 1$: $x\left(  t\right)  =\int_{0}^{t}\xi\eff ^{-1}F_{0}\left(
\tau\right)  e^{-\xi\eff ^{-1}C\left(  t-\tau\right)  }d\tau$. After some
calculation, this leads to \cite{Remark}:
\begin{equation}
C\MT \left(  t\right)  =\frac{F_{0}^{2}}{\xi\eff ^{2}}\frac{2T_{r}T_{p}%
}{4T_{r}^{2}-T_{p}^{2}}\left(  \tfrac{2T_{r}}{T_{p}}e^{-\frac{2\left|
t\right|  }{T_{p}}}-e^{-\frac{\left|  t\right|  }{T_{r}}}\right)
\label{VcorrFTelegraph}%
\end{equation}
The two characteristic timescales are now the MT relaxation time
$T_{r}=\xi\eff /C$ and the motor processivity time $T_{p}$. Inserting
\eqr\ref{VcorrFTelegraph} into \eqr\ref{MSDstationary} gives a lengthy expression
which in the limit $t\gg\max(T_{p},T_{r},t\ssw )$ can be used to compute the
long-time diffusion constant%
\begin{equation}
D=\frac{{F_0}^{2}}{C^{2}}\frac{2K\left(  1-K\right)  t\ssw T_{p}}{\left(
2T_{r}+t\ssw \right)  \left(  t\ssw +T_{p}\right)  \left(  T_{p}%
+2T_{r}\right)  }\allowbreak\label{DConstantTelegraph}%
\end{equation}
As in the previous example (periodic MT oscillation), the vesicle
mobility is always maximized for $K\opt =1/2$, \ie  for an
intermediate binding strength. The optimal vesicle switching time $t\ssw \opt %
=\sqrt{2T_{r}T_{p}}$ is also easily obtained from
\eqr\ref{DConstantTelegraph}. Interestingly, for fixed $T_{r}$,
\eqr\ref{DConstantTelegraph} predicts the optimal ratio
$T_{p}/T_{r}=2$. For those values we obtain the maximal diffusion
constant $D\smax =F_{0}^{2}\left( \xi+F_{0}/v_{0}\right)
^{-1}C^{-1}/32$, \ie  stronger and faster motors ($F_{0}$ and
$v_{0}$ large), weaker MT confinement ($C$ \ small) and smaller MT
friction give rise to more efficient transport.

\textit{Multiple MTs. \ }For $n$ different oscillating MTs, the
particle velocity is $v\left(  t\right)  =\sum_{k=1}^{n}%
\delta_{k,B\left(  t\right)  }v_{MT_{k}}\left(  t\right)  $ with
$v_{MT_{k}}$ the velocity of the $k$th MT, $\delta_{k,l}$ the
Kroneker-delta and $B\left( t\right)  \in\left\{
0,1,2,..,n\right\}  $ the ``binding variable'' which indicates to
which MT the vesicle is bound at time $t$ ($0$ represents the
unbound state). For the two-MT situation in \figr1a, $n=2$ and
$v_{MT_{1}}=-v_{MT_{2}}$ (anticorrelated MT velocities). Using
\eqr \ref{VcorrFTelegraph} as before yields the resonance
condition: $k\on \gg k\off ,$ $k\off =\sqrt{2}\left(
T_{r}T_{p}\right)  ^{-1/2}$ and $T_{p}=2T_{r}$, which differs
slightly from the single MT case because it is now more favorable
to jump between the two MTs than to spend time in the unbound
state. The corresponding diffusion constant $D\smax =$
$F_{0}^{2}\left(  \xi+F_{0}/v_{0}\right)  ^{-1}C^{-1}/8$ is 4
times larger than in the single MT case. This indicates that the
efficiency of transport, as well as the natural strategy for
optimizing it, can depend on the effective number of participating
MTs.

\textit{Hitchhiking vs thermal diffusion.} In the limit of motor forces larger
than viscous forces ($F_{0}\gg \xi v_{0}$), we obtain the rough estimate $D_{\max
}\approx\frac{1}{8}\left(  F_{0}/C\right)  v_{0}$. Typical MT sliding
velocities $v_{0}\approx1$--$5\,\micron /\second$ and oscillation amplitude
$F_{0}/C\approx0.1$--$1\micron $ yield $D\smax
\approx0.01$--$0.5\micron^{2}\second^{-1}$. In comparison, a typical organelle with diameter 500nm
experiences a large cellular viscosity $\eta\approx0.3\,\pascal \cdot
\second$
\cite{Kinesin/Dynein} and has a thermal diffusion constant $D\sth %
\approx3\cdot10^{-3}\micron ^{2}\second^{-1}$. On the other hand,
nanometer sized molecules
experience a much smaller effective viscosity (close to that of water
$\eta\approx10^{-3}\,\pascal \cdot \second$) and so have $D\sth
\approx10$--$100\,\micron ^{2}\second^{-1}$.
Therefore large objects like vesicles and organelles can strongly benefit from
hitchhiking while smaller molecules are more efficiently transported by
thermal diffusion.

\textit{Hydrodynamic stress  and biased hitchhiking. } It is
straightforward to show that the hitchhiking process as described
above for large $t$ becomes an unbiased diffusive process even for
an asymmetric MT shaking \cite{InPrep}. We consider next an
interesting generalization of the hitchhiking model which gives
rise to biased directed transport.

In the following, we drop the previous assumption $\left\langle
Bv\MT \right\rangle =\left\langle B\right\rangle \left\langle v\MT
\right\rangle $ \ie  the statistical independence of the MT
velocity and the vesicle--MT binding process. For example, a
statistical coupling of binding and MT sliding can  appear when
the off rate $k\off \left(  v\MT \right)  $ becomes velocity
dependent due to a hydrodynamic drag force (with respect to the
essentially stationary cytoplasm) acting on a moving vesicle. The
presence of this force can break the symmetry of transport in two
different ways. The general form of the off rate will be $k\off
\left(  v\MT \right)  /k\off \left(  0\right) \approx1+c_{2}v\MT
^{2}+c_{3}v\MT ^{3}+\cdots$. The linear term must vanish, if we
make the plausible requirement that $k\off \left( v\MT \right)  $
is minimal for $v\MT =0$ (no hydrodynamic stress = maximal binding
strength). We assume that the two remaining terms $c_{2}v\MT ^{2}$
and $c_{3}v\MT ^{3}$ are small. We call the coefficient $c_{2}$
the \textbf{dynamical bias coefficient}, because it gives rise to
a particle drift only if the MT has different forward and backward
velocity. We call $c_{3}$ the \textbf{polarity bias}, as it can
give rise to drift even for a time reversal symmetric $v\MT \left(
t\right)  $. Physically, this term stems from the polarity of MT
surface and the resulting polarity of the interaction with the
vesicle.

We next derive the mean drift velocity\ $\left\langle v\left(
t\right) \right\rangle $ of the vesicle in the simplifying
limiting case of rapid binding equilibration \ie $k\on +k\off \gg
\omega_{\rm MT,char}$ where $\omega_{\rm MT,char}$ denotes the
characteristic oscillation frequency of the MT. In this limit, the
vesicle binding state equilibrates at each instant of time and the
conditional binding probability becomes a function of the
instantaneous MT velocity: $p\left(  B=1|v\MT \right) =k\on
/\left( k\off \left(  v\MT \right)  +k\on \right) \approx
K_{0}-K_{0}\left( 1-K_{0}\right) (c_{2}v\MT ^{2}+c_{3}v\MT ^{3})$
with $K_{0}=k\on /\left( k\off \left(  0\right)  +k\on \right)  .$
Consequently the vesicle's mean drift velocity simplifies to
$\left\langle v\left( t\right)  \right\rangle =\left\langle v\MT
\left(  t\right) p\left(  B=1|v\MT \left( t\right) \right)
\right\rangle =\left\langle v\MT \left( t\right) \right\rangle
K_{0}-K_{0}\left(  1-K_{0}\right) (c_{2}\left\langle v\MT
^{3}\left( t\right)  \right\rangle +c_{3}\left\langle v\MT
^{4}\left(  t\right)
\right\rangle $ and its long time mean (vesicle drift) reads:%
\begin{equation}
\overline{v}=-K_{0}\left(  1-K_{0}\right)  (c_{2}\overline{\left\langle
v\MT ^{3}\right\rangle }+c_{3}\overline{\left\langle v\MT ^{4}\right\rangle
}) \label{MeanDrift}%
\end{equation}
where the long time average $\overline{f}%
\equiv\lim_{T\rightarrow\infty}T^{-1}\int_{0}^{T}f\left(  t\right)
dt$. $\overline{\left\langle v\MT \right\rangle }$ vanishes
because of MT confinement.

As an example, we consider a (deterministic) asymmetric square
wave MT oscillation with $v\MT \left(  t\right)  $ periodically switching
between $v\MT =V_{1}$ for a time $T_{1}$ and $v\MT =V_{2}$ for time $T_{2}$
with $V_{1}V_{2}<0$ and $T_{1}V_{1}+T_{2}V_{2}=0$ (zero mean). For such a
choice from \eqr \ref{MeanDrift} we obtain $\overline{v}=-K_{0}\left(
1-K_{0}\right)  \left(  T_{1}+T_{2}\right)  ^{-1}(c_{2}\left(  T_{1}V_{1}%
^{3}+T_{2}V_{2}^{3}\right)  +c_{3}\left(  T_{1}V_{1}^{4}+T_{2}V_{2}%
^{4}\right)  )$ which in general has both the dynamical ($c_{2}...$) and the
polar ($c_{3}...$) drift contributions. For a completely symmetric shaking
($V_{2}=-V_{1},$ $T_{1}=T_{2}$), the dynamical part vanishes as
expected, whereas
the polar part always stays present with a sign opposite to $c_{3}$.

In conclusion, we have outlined a transport mechanism alternative
to the standard ``cargo hauled by a motor'' model.  Active
longitudinal motions of MTs, combined with even weak non-specific
cargo--MT binding, naturally lead to this transport scenario.
Remarkably, even small scale (spatially confined) MT oscillations
induce a long distance transport on the cellular scale.
Hitchhiking on approximately symmetric MT bundle arrangements as
in \figr1a inherently bears the feature of velocity and run-length
distribution symmetry often observed in BDT
\cite{BiDirectional,BiDirectional2}. Very large moment velocities
\cite{Amos,BiDirectional2,Orokos} (up to $>10\,\micron
/\second$)---another characteristic of BDT---may be attributable
to the enhanced motor efficiency known for cooperatively
synchronized motor batteries performing filament sliding
\cite{Satir} and elastic MT relaxations \cite{InPrep}. Other
signature behaviors of our mechanism, such as visible sliding of
MTs, tandem motion of vesicles, exponential velocity relaxations
and strong MT bending deformations have been observed recently
\cite{Orokos,InPrep}. In general, we expect a combination of
hitchhiking and hauling mechanisms to be present in varying
proportions for different cargos and cells. Future experiments
might reveal the relative importance of their contributions and
the ``master plan'' behind them.

We thank J. Crocker, B. Hoffman, P. Janmey, and P. Baas for
discussion. We are indebted to P.R. Selvin, V. Gelfand, B. Blehm,
H. Kim, and C. Kural for fruitful discussions and for sharing
their movies prior to publication. This work was supported in part
by NSF Grants DMR04-04674 and DMR04--25780.

\end{document}